\documentclass{article}

\usepackage{arxiv}

\usepackage[utf8]{inputenc} 
\usepackage[T1]{fontenc}    
\usepackage{hyperref}       
\usepackage{url}            
\usepackage{booktabs}       
\usepackage{amsfonts}       
\usepackage{nicefrac}       
\usepackage{microtype}      
\usepackage{cleveref}       
\usepackage{lipsum}         
\usepackage{graphicx}
\usepackage{natbib}
\usepackage{doi}
\usepackage{ulem}
\usepackage[english]{babel}
\usepackage[T1]{fontenc}
\usepackage[utf8]{inputenc}

\title{A response matrix determined with a coincidence-based acquisition for correction of
charge sharing spectral distortions in energy-resolved photon counting detectors}

\newif\ifuniqueAffiliation

\ifuniqueAffiliation 
\author{ \href{https://orcid.org/0000-0002-3617-2432}{\includegraphics[scale=0.06]{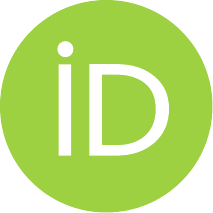}\hspace{1mm}Vincenzo Monaco}
	Universit`a di Torino and INFN sezione di Torino\\
	Via Pietro Giuria 1\\
	10125 Torino, Italy \\
	\texttt{vincenzo.monaco@to.infn.it} \\
}
\else
\usepackage{authblk}

\newbox{\orcid}\sbox{\orcid}{\includegraphics[scale=0.06]{orcid.pdf}}
\author[1,2]{
	\href{https://orcid.org/0000-0002-3617-2432}{\usebox{\orcid}\hspace{1mm}Vincenzo Monaco\thanks{\texttt{vincenzo.monaco@unito.it}}}%
}
\author[3,4]{
	{\hspace{1mm}{Luca Brombal}}
}
\author[5,6]{
	{\hspace{1mm}Pasquale Delogu}
}
\author[7,6]{
	{\hspace{1mm}Alessandro Feruglio}
}
\author[8,9]{
	{\hspace{1mm}{Massimiliano Fiorini}}
}
\author[3,4]{
	{\hspace{1mm}{Renata Longo}}
}
\author[1]{
	{\hspace{1mm}Luca Marchetti}
}
\author[1]{
	{\hspace{1mm}Anna Maria Poli}
}
\author[3,4]{
	{\hspace{1mm}{Luigi Rigon}}
}
\author[7,6]{
	{\hspace{1mm}{Valeria Rosso}}
}

\affil[1]{Dipartimento di Fisica, Universit\`a degli Studi di Torino, Via Pietro Giuria 1, Torino, 10125, Italy}
\affil[2]{INFN Torino, Via Pietro Giuria 1, Torino, 10125, Italy}
\affil[3]{Dipartimento di Fisica, Universit\`a degli Studi di Trieste,
Via Valerio 2, Trieste, 34127, Italy}
\affil[4]{INFN Trieste, Via Valerio 2, Trieste, 34127, Italy}
\affil[5]{Dipartimento di Scienze Fisiche, della Terra e dell'Ambiente, Universit\`a di Siena,
Via Roma 56, Siena, 53100, Italy}
\affil[6]{INFN Pisa, Largo Bruno Pontecorvo 3, Pisa, 56127, Italy}
\affil[7]{Dipartimento di Fisica, Universit\`a di Pisa, Largo Bruno Pontecorvo 3, Pisa, 56127, Italy}
\affil[8]{Dipartimento di Fisica e Scienze della Terra, Universit\`a degli Studi di Ferrara,
Via Saragat 1, Ferrara, 44122, Italy}
\affil[9]{INFN Ferrara, Via Saragat 1, Ferrara, 44122, Italy}
\fi

\begin{document}
\maketitle

\begin{abstract}
	A model-independent method is proposed to characterize and correct charge sharing spectral distortions in energy-resolved X-ray acquisitions with pixellated photon-counting detectors.
	The technique is based on the determination of a coincidence-based response matrix (CBRM) through a preliminary calibration with a uniform irradiation and an arbitrary
	polychromatic spectrum.
	The calibration requires the collection of the number of coincidences between a reference pixel and its neighbours for different combinations of energy bins,
	in order to calculate a set of charge sharing probabilities which are independent of the input spectrum.
	A detector response matrix is determined, which can afterwards be applied to correct other spectra acquired with the same detector and a conventional multi-comparator electronics,
	without introducing penalties in terms of processing time.
	The technique was validated with Geant4 simulations of a 1 mm thick CdTe detector and with data collected with a pixel hybrid detector
	made of a 300 {\textmu}m thick silicon sensor coupled to a Timepix4 ASIC chip.
	The differences between the reconstructed spectra and reference distributions of an ideal detector or a 3x3 offline clustering algorithm were evaluated in terms of mean absolute percentage errors.
	It is demonstrated that the response matrix can restore the spectral information with a performance close to standard clustering algorithms and
	is less affected by noise artifacts than analog charge summing techniques.
\end{abstract}


\section{Introduction}
\label{intro}

Photon-counting detectors (PCDs) are rapidly evolving as a new technology for X-ray imaging
in medical diagnostics \citep{Taguchi2013, Iwanczyk2015, Willemink2018}, material analysis \citep{Zemlicka_2011}, security inspection \citep{CHEN2023167886}
and other applications.
Planar semiconductors, mainly high-Z cadmium-telluride (CdTe) and cadmium-zinc-telluride
(CZT), provide a direct conversion of the released energy in a measurable charge \citep{rossi2010}. When coupled to electronics with pulse-height capabilities,
it is possible to detect each interacting photon and
measure its energy.

In X-ray Computed Tomography, PCDs have demonstrated suppression of electronic noise, higher spatial resolution,
reduced dose for the same image quality, higher dynamic range,
improved elemental decomposition and enhanced contrast agent discrimination
with respect to traditional charge-integrating detectors \citep{flohr2020}.
A first commercial scanner for X-ray photon counting Computed Tomography was introduced by Siemens Healthineers in 2021, confirming the
clinical benefits of this technology \citep{doi:10.1148/radiol.212579} \citep{Cademartini2023}.

Despite their intrinsic advantages, the performance of PCDs in spectroscopic X-ray imaging is limited by two deleterious effects:
the sharing of the signals between neighbouring pixels \citep{Taguchi2016, Tanguay2018} and the overlap of pulses from events occurring
too close in time (pulse pileup, PPU) \citep{Knoll2010xta, Wang2011}.
Both these effects distort the spectral distribution of counts in each pixel and degrade the information obtainable from the reconstructed
images \citep{Roessl2011CombinedEO}.
Pileup distortions are particularly critical in clinical photon counting Computed Tomography, where the X-ray flux could be
of the order of $10^9 \mathrm{s}^{-1} \mathrm{mm}^{-2}$.

The sharing of the charge between different pixels is due to the induction of signals on two or more nearby pixels while the carriers' charges
migrate toward the electrodes \citep{KIM2011233}.
This effect is due to the size of the charge cloud and to the superposition of the weighting potentials of adjacent pixels. It
becomes significant in high spatial resolution applications, where the pixel pitch is small with respect to the sensor thickness.
Other spectrum distortions in high-Z sensors arise from the emission of characteristic fluorescence photons or Compton photons which can
interact in other detector pixels or exit from the detector.
In the following, we use a wide definition of charge sharing (CS) which includes all the mechanisms that could generate the splitting
of the charge produced by an interacting photon between different pixels, giving rise to multiple hits with lower charge.
Charge sharing distortions can be attenuated by using pixels of larger dimensions, but this comes at the cost of both reduced spatial resolution and a worsening of pileup
effects which, on the contrary, are reduced with smaller pixels.

The readout electronics has a fundamental role in optimizing the performance of PCD devices.
A review of electronics developed for spectroscopic photon counting imaging can be found in \citep{Ballabriga_2016}.
The typical technique for charge sharing compensation is based on the on-line clustering of the energy measured by nearby pixels \citep{Maj_2012, Koenig2013}.
A dedicated analog processing is adopted to sum the energies released in groups of pixels and inter-pixel comparators are employed to assign the resulting
value to the pixel with the largest energy release.
These "analog charge summing" (ACS) algorithms, already implemented in the electronics of the Medipix family \citep{Ballabriga_2013}
and in other readout circuits \citep{veale2011, Ullberg2013, Bellazzini_2015, 7027239, DITRAPANI2020163220}, are very effective in compensating charge sharing effects,
but introduce an additional dead-time which worsens the effect of pileup at high input rates \citep{10.1117/12.2293591}.

Other approaches for the correction of PPU and/or CS effects are based on the post-processing of the acquired spectral images.
Most of the proposed analytical methods make use of semi-empirical response functions based on a model of the processes producing
spectral distortions, with parameters determined from simulations or data fitting \citep{Ding2012, Cammin2016, Dreier2018}.
In most of these studies the response function is defined for a particular irradiation setup and detector configuration, and
it is not guaranteed to be of general use in image reconstructions with different conditions.

In recent years, multi-energy inter-pixel coincidence counters (MEICC) were proposed for charge sharing corrections \citep{Taguchi2020}.
This method requires a dedicated readout electronics to count the number of coincidences between different energy bins for each pixel and its neighbours.
Many coincidence gates and counters (of the order of $N^2$ per pixel where $N$ is the number of energy bins) are needed.
Post-processing algorithms of the data collected with MEICC counters were proposed in \citep{10.1117/12.2654389},
and with a deep-learning approach in \citep{Zhao_2024}, showing the promising potentiality of the method for compensation of CS effects.

In this paper a statistical post-processing method for the characterization and correction of charge sharing distortions of PCD X-ray spectra is proposed.
The method is based on the collection of calibration data with a reference polychromatic spectrum and a uniform illumination of the detector to determine
the elements of a spectral response matrix. At this stage it is required that the number of coincidences between the signal of one pixel and the
analog sum of its neighbouring pixels are counted for all the possible combinations of energy bins.
Assuming an equilised behaviour of all the PCD pixels, the calibration requires a simple coincidence circuit involving only a limited number of pixels.
The calibration allows to define the probabilities that the initial energy is split
between one pixel and its neighbours.
These probabilities are used to define the elements of a matrix, named “coincidence-based response matrix” (CBRM), which inversion allows to retrieve the
input energy distribution in the absence of charge sharing.
The CBRM is independent of the input spectrum and can therefore be used to correct CS distortions in subsequent acquisitions
with different spectra. The correction requires only the integral number of counts collected for different energy bins by each pixel.
Therefore, once the response matrix is determined, the image acquisitions can be performed with conventional multi-comparator readout electronics,
without introducing additional dead-times. Compared to other semi-empirical post-processing algorithms, the method is based only on data inputs,
is independent of models or simulations, is simple and of general applicability.

The method was validated with Monte-Carlo simulations of a CdTe detector and with data acquired with a silicon pixel sensor coupled with the
Timepix4 \citep{Llopart_2022} readout electronics using monochromatic and polychromatic X-ray beams.

\section{Methods}
\label{method}

\subsection{Acquisition system}
\label{method:daq}

In most of the electronics for spectroscopic photon counting the signal from each pixel, after an amplification and shaping stage,
is fed to a multiple-comparator system comprising a bank of discriminators with different thresholds.
Each comparator (henceforth identified by an index $i$) is associated to a counter register, which is
incremented every time the rising part of the pixel pulse crosses the corresponding threshold.
In the following, we assume that the thresholds are expressed in terms of energy values.
We indicate with $N_{i}$ the number of events for which the energy released by a photon in one pixel is above the
threshold $i$ and with $N_{i}'$ the number of raw counts for counter $i$.
In general the two values are different due to charge sharing, pileup, electronic noise,
detector resolution and other effects. If pileup effects are negligible, the relation between the number of raw
counts $N_i'$ above threshold and the number of counts $n_i'$ in the energy bin $i$ defined by two adjacent thresholds is

\begin{equation}
n_i'=N_i'-N_{i+1}'
\label{eq0}
\end{equation}
The same relation holds for the number of input events.

\begin{figure}[ht]
\centering
\includegraphics[width=0.95\textwidth]{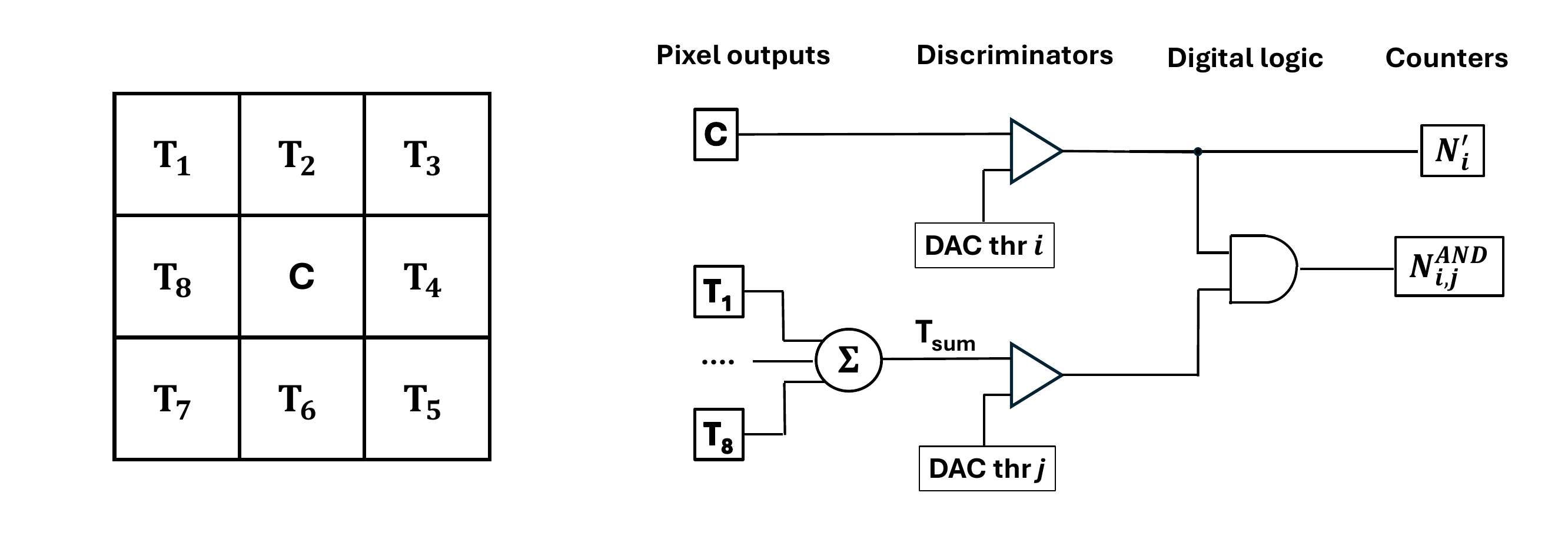}
\caption{Left: labelling of pixels in a 3x3 block. Right: scheme of the acquisition system used to store the information needed
to determine the elements of the spectral response function of the detection system. The signals from a reference pixel (C)
and the sum of the analog signals from the 8 surrounding pixels ($T_{sum}$), after amplification and shaping,
are compared against a pair of energy thresholds $i$ and $j$.
The number of threshold crossings for the reference pixels and for the logical AND are stored in counters.
The procedure is repeated for all the possible combinations of threshold values.
}
\label{fig1}
\end{figure}

In this study, the response matrix is determined by counting the number of correlated events between a reference pixel
and the 8 nearest neighbours. The labelling defined in Fig.\ref{fig1}(left) is used, where C is the pixel under study
(in the following called "reference pixel"), and the symbols $T_l$ with $l=1,...8$ identify the 8 pixels surrounding the reference one.
The proposed readout scheme for the calibration phase is schematized in Fig.\ref{fig1}(right).
The analog sum of the signals from the 8 peripheral pixels ($T_{sum}$) and the signal from the pixel C are discriminated
against two programmable thresholds. The two thresholds can be selected from a set of values defining a fine division of the
energy range in $L$ bins. A minimum noise discrimination threshold is applied to avoid fake counts.

For each pair of possible thresholds $i$ and $j$ for the C and T pixels, the number of counts from the reference pixel $N_{i}'$
and from the logical AND combination of the two discriminator outputs $N^{AND}_{i,j}$ are saved in counters.
If pileup effects are negligible, the relation between the number of coincident events $n^{AND}_{i,j}$ in two energy bins $(i,j)$
and the number of coincidences above two thresholds $N^{AND}_{i,j}$ is given by:

\begin{equation}
	n^{AND}_{i,j}=N^{AND}_{i,j}-N^{AND}_{i+1,j}-N^{AND}_{i,j+1}+N^{AND}_{i+1,j+1}
	\label{eq1}
\end{equation}

The calibration is performed with a uniform irradiation of the detector and different acquisitions for all the possible
combinations of the two threshold levels. Assuming a detector with equalised pixel outputs, the calibration can be performed
using only one coincidence logic involving a single group of 8+1 pixels.

\subsection{Determination and application of the CBRM matrix}
\label{method:model}

\subsubsection{Charge sharing model}
\label{method:model:cs_model}

We assume that the useful energy range is divided into $L$ bins of equal width $\Delta E$.
The model could be extended to a set of energy bins of
different widths but, for sake of simplicity, the study was conducted considering only bins
of equal width. In the following we will indicate with an index $i$ a count or an event
in the energy bin between $i\Delta E$ and $(i+1)\Delta E$.
We focus on a given pixel of the detector, identified as the reference pixel C of Fig.\ref{fig1}(left).
The charge sharing model used in this study is based on the following assumptions:

\begin{itemize}
	\item When the incoming photon interacts in a pixel, the released energy can be shared only with the eight surrounding pixels.
	\item The charge sharing is modelled by a set of probabilities $p^{(out)}_{i,j}$ that an event in the reference pixel C with input energy at bin $k$
	 is split into multiple lower energy hits, with an energy $j$ in the sum of the 8 surronding pixels and with energy in the bin $i=k-j$
	in the reference pixel. We refer to these events as "CS-OUT" transitions.
	\item Whenever the interaction occurs in one of the 8 pixels $T_{l}$ around C with deposited energy at bin $k=i+j$,
	the probability that an energy at bin $i$ is detected in the reference pixel is $p^{(in)}_{j,i}$.
	These events are named "CS-IN" transitions.
	In the calibration phase it is also assumed that, when an interaction occurs in a peripheral pixel and a coincidence
	with the central pixel is detected, the cluster of pixels where the released energy is split does not extend
	outside the 3x3 block defined in Fig.\ref{fig1}(left). The residual energy $j$ is therefore collected
	by the 8 peripheral pixels.
	\item For a flat-field calibration irradiation, the input spectrum and the energy distribution of counts is the same for all the pixels.
	The detector response is assumed to be uniform among the pixels.
\end{itemize}

\begin{figure}[ht]
	\centering
	\includegraphics[width=0.40\textwidth]{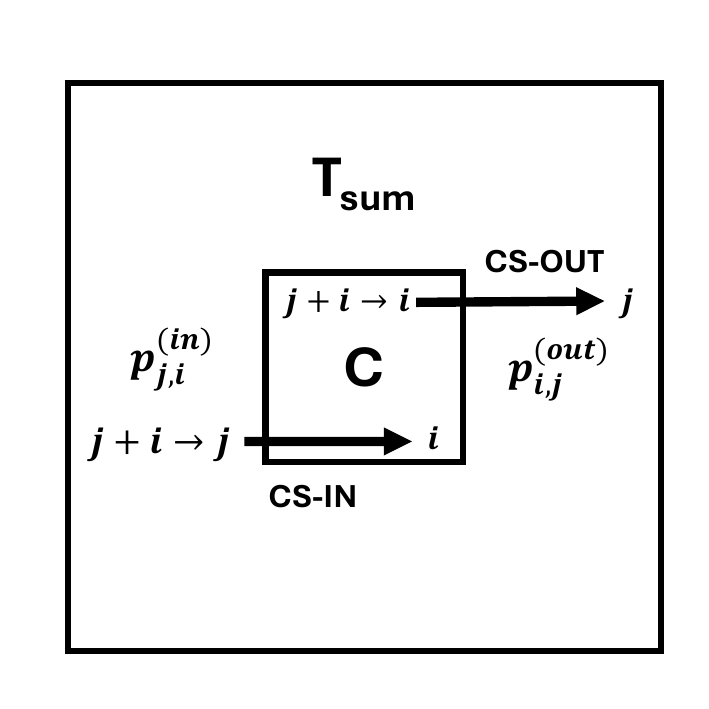}
	\caption{Schematic of the sharing probabilities used in the model. If the interaction occurs in the reference pixel C,
	the released energy can be split into an energy at bin $i$ in the reference pixel and an energy at bin $j$ distributed among the surrounding
	8 pixels with probability $p_{i,j}^{(out)}$ (CS-OUT transitions).
	In a CS-IN transition the interaction event occurs in one of the 8 pixels around the reference one, and the deposited energy
	is split into a signal at bin $i$ of the reference pixel and a signal at bin $j$ of the peripheral pixels with probability
	$p^{(in)}_{j,i}$. In both cases the input energy is at bin $k=i+j$.}
	\label{fig2}
\end{figure}

The definition of charge sharing CS-IN and CS-OUT probabilities is schematized in Fig.\ref{fig2}.

The charge sharing produces a distortion in the measured number of counts in energy bin $i$ of a
reference pixel due to three possible processes listed below.

\begin{enumerate}
	\item The energy released by a photon interaction in the reference pixel with energy at bin $k$ is shared with the nearby pixels,
	reducing the energy measured in the reference pixel to bin $i$ and adding a count to the corresponding energy bin.
	The probability of this CS-OUT transition is $p^{(out)}_{i,k-i}$, with $k-i$ the residual energy released in the peripheral pixels.
	\item An event in the reference pixel with deposited energy $i$ is shared with the neighbouring pixels,
	reducing the number of counts at bin $i$. This CS-OUT transition has probability $p^{(out)}_{j,i-j}$,
	where $j$ is the residual energy in the reference pixel and $i-j$ is the energy released in the nearby pixels.
	\item An interaction event occurring in one of the eight neighbouring pixels with energy at bin $k$
	is shared with the reference pixel, adding a count at bin $i$ of the reference pixel and reducing the
	energy of the peripheral pixels to $k-i$. The probability of this CS-IN transition is $p^{(in)}_{k-i,i}$.
\end{enumerate}

For a generic number $L$ of energy bins, the spectral distortion due to charge sharing can be described in terms of
transition probabilities by the following equation:

\begin{equation}
n^{(c)'}_{i}
= n^{(c)}_{i}+\sum_{k=i+1}^{L-1} p^{(out)}_{i,k-i}n^{(c)}_{k}-\sum_{j=0}^{i-1} p^{(out)}_{j,i-j}n^{(c)}_{i}+ \sum_{k=i}^{L-1} p^{(in)}_{k-i,i}n^{(t)}_{k}
\label{eq2}
\end{equation}

where $n^{(c)'}_i$ is the number of raw counts at energy bin $i$ of the reference pixel C,
$n^{(c)}_{i}$ the number of input events at bin $i$ in the pixel C,
$n^{(t)}_k$ the number of input events at energy bin $k$ for the sum of the 8 pixels surrounding
the reference one. The three last terms in Eq.\ref{eq2} correspond to the three possible distortions of the number of counts at bin $i$ described above.

It has to be observed that, for an event occurring in one of the peripheral pixels, the probability to share a fraction of
its energy with the reference pixel is one eighth of the probability that an event occurring in the pixel C
with the same input energy shares the same fraction of energy with the surrounding pixels.
Therefore, the following relation holds between the probabilities of CS-OUT transitions $p^{(out)}_{i,j}$ and of
CS-IN transitions $p^{(in)}_{j,i}$:

\begin{equation}
	p^{(out)}_{i,j}=8p^{(in)}_{j,i}=q_{i,j}
	\label{eq3}
\end{equation}

Eq.\ref{eq2} can thus be expressed in terms of the $q_{i,j}$ probabilities of Eq.\ref{eq3} as:

\begin{equation}
	n^{(c)'}_{i}=n^{(c)}_{i}+\sum_{k=i}^{L-1} q_{i,k-i}\left(n^{(c)}_{k}+\frac{1}{8}n^{(t)}_{k}\right)-\sum_{j=0}^{i} q_{j,i-j}n^{(c)}_{i}
	\label{eq4}
\end{equation}

\subsubsection{CBRM matrix and spectral corrections}
\label{method:model:cbrm_calc}

The acquisition system of Fig.\ref{fig1}(right) is used to collect the number of raw counts $n_{i}^{'}$ for one pixel
and of coincidences $n^{AND}_{i,j}$ between this pixel and its 8 neighbours, during a uniform illumination
of the whole detector with a polychromatic X-ray beam.
The number of coincidences can be expressed in terms of the transition probabilities $q_{i,j}$,
defined by Eq.\ref{eq3}, and of the number of input events in one pixel $n_{i}^{(c)}$ and in the 8 neighbours $n_{i}^{(t)}$, as:

\begin{equation}
	n^{AND}_{i,j} = \left( p^{(out)}_{i,j}n^{(c)}_{i+j}+p^{(in)}_{j,i}n^{(t)}_{i+j}\right) = q_{i,j} \left( n^{(c)}_{i+j} + \frac{1}{8}n^{(t)}_{i+j}\right)
	\label{eq5}
\end{equation}

For a uniform illumination, all the pixels have the same energy distribution, and therefore:

\begin{equation}
	n^{(t)}_i = 8 n^{(c)}_{i} = 8 n_i
	\label{eq6}
\end{equation}

In this case, the probabilities $q_{i,j}$ in Eq.\ref{eq5} are given by:

\begin{equation}
	q_{i,j} = \frac{n^{AND}_{i,j}}{2n_{i+j}}
	\label{eq7}
\end{equation}
where $n_{i+j}$ in the denominator is the number of input events at energy bin $i+j$.

Substituting Eq.\ref{eq7} in Eq.\ref{eq4}, the number of raw counts in the energy bin $i$ of one pixel can be expressed as;

\begin{equation}
	n_i'=n_i+\sum_{j=i}^{L-1} n_{i,j-i}^{AND}-\frac{1}{2}\sum_{j=0}^{i} n^{AND}_{j,i-j}
	\label{eq8}
\end{equation}

Eq.\ref{eq8} allows to determine the number of input events $n_i$ from the number of raw counts $n_{i}'$ and of coincidences collected in the calibration run.
Therefore, the transitions probabilities $q_{i,j}$ can be determined from Eq.\ref{eq7}.

Once the transition probabilities are determined with a concidence-based calibration, they can be applied to compensate charge sharing distortions
for different spectra acquired with a
readout based on a bank of discriminators and counters. If the set of thresholds is the same employed in the calibration,
 the number of input events $n_i$  can be retrieved from the number of raw counts $n_i'$ by solving equation Eq. \ref{eq4}
  which, for a uniform illumination, becomes:

  \begin{equation}
	n_i'= \left[ 1 +q_{i,0}-\sum_{j=0}^{i-1} q_{j,i-j}\right] n_i + 2\sum_{j=i+1}^{L-1} q_{i,j-i} n_{j} =\sum_{j=0}^{L-1} A_{i,j} n_{j}
	\label{eq9}
\end{equation}

The factors $A_{i,j}$ in Eq.\ref{eq9} are the elements of the CBRM matrix.
The matrix is triangular and can be easily inverted to retrieve the original number of events in each energy bin.
In practice, Eq.\ref{eq9} can be solved in an iterative way, starting from the highest energy bin and proceeding toward lower energy bins.
The solution for the number of input events at bin $i$ is:

\begin{equation}
	n_i = \frac{n_i'-2\sum_{j=i+1}^{L-1}q_{i,j-i}n_{j} }{1+q_{i,0}-\sum_{j=0}^{i-1}q_{j,i-j}}
	\label{eq9c}
\end{equation}

\subsubsection{Reduction of the CBRM matrix}
\label{mehod:model:matrix_reduction}

The existing electronics includes a relatively small number of discriminators,
in general lower than the number of energy bins needed for an accurate definition of the response matrix.
Therefore, the transition probabilities collected in the calibration run must be scaled to match
the number of energy bins of the standard readout.
For simplicity we assume that in the calibration run and in a standard acquisition the same energy
range is divided respectively in $L'$ and $L$ energy bins, with $L'$ a multiple of $L$ ($L'=W\cdot L)$.
We indicate with $i$ the energy bins of the standard acquisition and with $i'$ those of the calibration run,
with $0\le i<L$ and $0\le i'<L'$.
The number of input events and measured counts in the energy bin $i$ for the standard acquisition are $m_{i}$ and $m_{i}'$, respectively.

For wide energy bins the hypothesis that the sum of the mean energies at bins $i$ and $j$ in the reference pixel
and in the 8 neighbour pixels is equal to the mean energy of bin $i+j$ does not always hold.
The spectral distortions are thus parameterized in terms of three-fold CS-OUT probabilities $P^{(out)}_{k,i,j}$
that an event with initial energy $k$ released in the reference pixel is split into a count at bin $i$ in the
same pixel and a count at a bin $j$ in the 8 surrounding pixels, with $k$ not necessarily equal to $i+j$.
If the set of thresholds employed in the standard acquisition is the same of the calibration run ($W=1$)
the three-fold probabilities coincide with those defined in the calibration $P^{(out)}_{k,i,j}=p^{(out)}_{i,j}$,
with $k=i+j$ for the assumption made in the charge sharing model to define the transition probabilities for energy bins of small width.

In case $W$ is an integer greater than 1, the probability $P^{(out)}_{k,i,j}$ can be estimated from
the transition probabilities $q_{i,j}$ determined in the calibration run as:

\begin{equation}
	P^{(out)}_{k,i,j}=\frac{1}{W}{\left| \sum_{i'=iW}^{(i+1)W-1} \sum_{j'=jW}^{(j+1)W-1} q_{i',j'} \right|}_{(i'+j')=kW}
	\label{eq10}
\end{equation}

The factor $W$ at the denominator of Eq.\ref{eq10} takes into account that, assuming a uniform
energy distribution inside an energy bin, the probabilities $q_{i,j}$ apply to a number of counts a factor
$W$ lower than the number of counts in the wider bin defined for the standard acquisition.
Similarly, probabilities for CS-IN transitions $P^{(in)}_{k,j,i}$ can be defined, where the indexes refer
to the input energy bin ($k$) and to the energies shared in the peripheral pixels ($j$) and in the reference pixel ($i$).

The same relation of Eq.\ref{eq3} holds between the CS-IN and CS-OUT transitions:
\begin{equation}
	P^{(in)}_{k,j,i} = \frac{1}{8} P^{(out)}_{k,i,j} = \frac{1}{8} P_{k,i,j}
	\label{eq11}
\end{equation}

The charge sharing probability that an event with initial energy $k$ in the reference pixel produces a count at
energy $i$ in the same pixel, regardless the energy shared with the nearby pixels, is given by:

\begin{equation}
	Q_{k,i}=\sum_{j=0}^{k} P_{k,i,j}
	\label{eq12}
\end{equation}

The number of raw counts $m_i^{(c)'}$ in a reference pixel for a generic acquisition with a limited number of energy bins
can be expressed as a function of the input number of events in the reference pixel ($m_i^{(c)}$) and in the 8 neighbouring pixels
($m_i^{(t)}$) using the transition probabilities defined in Eq.\ref{eq12}, as:

\begin{equation}
m_{i}^{(c)'}=m_{i}^{(c)}+\sum_{k=i+1}^{L-1} Q_{k,i} m_{k}^{(c)} - \sum_{k=0}^{i-1}Q_{i,k} m_{i}^{(c)} + \frac{1}{8}\sum_{k=i}^{L-1} Q_{k,i}m_{k}^{(t)}
\label{eq13}
\end{equation}

For a uniform illumination with $m_i^{(t)}=8m_{i}^{(c)}=8m_i$, Eq.\ref{eq13} becomes:

\begin{equation}
	m_{i}' = m_{i} + 2 \sum_{k=i+1}^{L-1} Q_{k,i} m_{k} +  Q_{i,i}m_{i} -\sum_{k=0}^{i-1} Q_{i,k}m_{i} = \sum_{k=0}^{L-1} A_{i,k}^{(r)} m_k
\label{eq14}
\end{equation}

Eq.\ref{eq14} defines the elements of a new $L\times L$ response matrix $A^{(r)}$, adapted to the wider energy bins of the multi-comparator system
with a limited number of energy thresholds. The inversion of the reduced response matrix, or the iterative solution of Eq.\ref{eq14}, allows to
retrieve the number of input events $m_{i}$.

\subsection{Simulation of a CdTe detector}
\label{method:simulation}

In order to validate the method described in the previous sections, the response of a pixelated CdTe detector of 1 mm thickness
and segmented in 9x9 pixels was simulated with the Geant4 toolkit \citep{AGOSTINELLI2003250} version 11.1.3. Different pixel sizes were considered, between 100 and 300 \textmu m.
X-ray photons
were generated with a direction perpendicular to the detector, and with an energy randomly selected from a polychromatic X-ray spectrum produced with the Spektr code \citep{Spektr}.
The photon source was placed at a fixed distance from the detector with transversal coordinates uniformly distributed to illuminate the whole detector.

The physics lists described in \citep{Tykhonov_2023} were used in Geant4 for an accurate simulation of emission and reabsorption of fluorescence and Compton
photons. The method described in the same paper was followed to simulate energy sharing due to the electron
charge cloud size. Each pixel is segmented in a matrix of 7x7 subpixels.
For an incoming photon impinging in a given sub-pixel, an interaction point is randomly selected in the subpixel area
with a uniform distribution. The released energy is therefore split into 500 equivalent parts, which are randomly distributed
across the detector area according to a 2D isotropic Gaussian distribution centered at the interaction point.
The energy in each pixel is the sum of the energy parts falling into the pixel area.
In our simulation the gaussian standard deviation was fixed to $\sigma=15$ \textmu m, assumed independent of the energy and of the
interaction depth. This value is consistent with those typically estimated in previous studies for a 1 mm CdTe converter
(an average value of 13 \textmu m was estimated in \citep{KOCHMEHRIN2020164241}, an initial cloud radius of $17.4\pm1.4$ \textmu m in \citep{VEALE2014218}).

Several factors were not included in this simple simulation, like detector resolution, charge cloud diffusion and trapping,
weighting field cross-talk, collection efficiency, signal formation, inter-pixel gaps, passive materials.
Noise effects were simulated by adding a random gaussian spread to the energy in each pixel.
Pulse shape and pileup effects were not included in this study.
It has to be remarked that this simple simulation was not intended to reproduce the real performance of a CdTe detection system,
but to characterize the effects of charge sharing mechanisms (including fluorescence and Compton emissions, reabsorption or escapes)
for the validation of the statistical correction method described in Sect.\ref{method:model}. Actually, the CBRM correction is not based on the
knowledge of the details of the detector response as it uses a CBRM estimation based on the collection of coincidence counts to characterize all the contributions
to charge sharing effects.

For each uniform illuminations and detector conditions $10^7$ events were generated, corresponding to $1.23\cdot 10^5$ input photons per pixel.
The generated events were divided into 10 data sets, and the analysis was repeated independently for each noise realization to evaluate the statistical
fluctuations of the results.
Only the central 5x5 pixels were analysed, excluding the edges of the detector.
To determine the response matrix, each of these pixels was correlated with the sum of the energies of its 8 nearest neighbours.

\subsection{Data taking with a Timepix4 hybrid detector}
\label{method:timepix}

The experimental validation of the proposed method was performed by analysing data collected by a hybrid detector consisting
of a 300 \textmu m thick p-on-n silicon detector with pixels of 55 \textmu m pitch bound-bonded to a Timepix4 ASIC chip.
 The Tipepix4 ASIC was developed by the Medipix collaboration and features $448 \times 512$ channels with individual charge amplifiers
 and thresholds programmable at pixel level with local DACs \citep{Llopart_2022, BALLABRIGA2023167489}. The total sensitive area is 24.64 mm $\times$ 28.16 mm.
 A data-driven readout modality provides, for each pixel whose signal crosses a threshold,
 the time of arrival (ToA) and the time the signal exceeds the threshold (time over threshold, ToT). The data were transmitted through two 2.56 Gbps links to an external readout system and saved on file.
 In the offline analysis the ToA was employed to identify clusters of simultaneous hits
 in neighbouring pixels while the ToT allowed, after proper calibration, to determine the energy deposited in each pixel. It has to be underlined that, even if the offline analysis
 was performed on data lists containing information for each pixel hit,
the correction method of the present study is intended to be applied to the integral number of counts stored internally in a chip, with obvious
advantages in terms of hardware simplicity and readout speed.

Monochromatic X-ray beams with 18 energies between 8.5 and 40 keV were produced at the SYRMEP beamline of the Elettra Synchroton in Trieste, Italy \citep{Elettra2010}.
 Details on the experimental conditions and on the calibration procedures followed to equalize the thresholds and convert the ToT measurements into energy values are
 described in \citep{FERUGLIO2024169844}. The SYRMEP beam cross-section, defined by tungsten slits, was 5.2 mm (vertical) $\times$ 28.6 mm (horizontal) with a vertical gaussian profile,
 covering 48k detector pixels. The detector was moved vertically during the acquisition to illuminate all the pixels.

Acquisitions with polychromatic spectra were performed at the PEPI Laboratory (Trieste, Italy) \citep{pepi_2023} with a microfocus X-ray tube with W anode operated at 50 kV and 0.1 mA.
The CBRM matrix was determined with a flat-field acquisition and afterwards applied to
correct the spectrum attenuated by a solutions of Ag contrast agent.
For these data sets, the detector calibration was based on test-pulses and on the energies of the K-edges
of different contrast agents (Ag, Ba, I). The analysis was restricted to a region-of-interest (ROI)
of 1500 pixels, corresponding to the central part of the vial containing the Ag solution.

The noise discrimination thresholds were set to 1000 e- (3.62 keV) for the SYRMEP acquisition, and to 300 e- (1.0 keV)
for the X-ray tube acquisition, both well above the Timepix4 typical equivalent noise charge (ENC) of 70 e- rms \citep{BALLABRIGA2023167489}.
In both tests the detector was operated at a bias voltage of 100 V.

Pixels hits with ToA values within an interval of 2 \textmu s were grouped to form an event.
The statistics collected at SYRMEP consisted in a number between 18 and 40 million of pixel hits for each beam energy while,
for the polychromatic acquisitions, 10M hits were collected for the calibration run and 4M for the Ag sample within the ROI.
The data sets were divided into 10 parts which were analysed independently for the evaluation of the statistical errors.

In all the acquisition the count rate per pixel was low enough to neglect the occurrence of pileup events.
The mean number of pixel hits per event was 2.5 and 1.5 respectively in the SYRMEP and X-ray tube acquisitions. Considering the area
covered by the beam or by the selected ROI, the probability that multiple hits in the same or adjacent pixels were produced by
 independent incident photons was estimated to be $<1$ per mill at SYRMEP and $<1 \%$ for the acquisitions with the X-ray tube.

\subsection{Validation of the CBRM correction}
\label{method:validation}

The spectra corrected with the CBRM matrix were compared with the energy distributions of an ideal detector and with other CS compensation techniques.

In the simulation studies, the CBRM corrections were compared with reference distributions obtained by summing the energies of all the pixels
for each event. The reference distributions were also compared with the simulation of a C8P1 charge summing algorithm \citep{7027239}
with summing nodes comprising 4 pixels. The same noise discrimination threshold was applied to the charges of individual pixels,
to the analog sums of the C8P1 algorithm,
 and to the sum of the charges from the 8 nearest neighbours used for the coincidence counts.

For the experimental data, the CBRM corrections were compared with the results of a clustering algorithm based on 3x3 summing nodes.
This clustering algorithm assigns to each pixel the sum of its energy and of the eight surrounding pixels and selects the clusters with the
maximum value of the sum. One or more non-overlapping 3x3 clusters are identified and, for each selected cluster, the energy
sum is assigned to the pixel with maximum individual energy.

The agreement between different spectra were quantified in terms of mean absolute percentage error (MAPE), defined as:

\begin{equation}
	MAPE = \frac{100}{m}\sum_{i=0}^{m-1} \frac{\mid n_{i}^{corr}-n_{i} \mid}{n_{i}}
	\label{eq_mape}
\end{equation}
where $m$ is the number of energy bins, $n_{i}^{corr}$ the number of counts in one pixel at energy bin $i$ after correction with the CBRM method or
obtained by the ACS algorithm,
and $n_{i}$ the number of counts in the same pixel at energy bin $i$ for the reference spectrum.
Bins with a number of counts lower than $10\%$ of the average number of bin counts were excluded from the sum in Eq.\ref{eq_mape} to limit the sensitivity
of the MAPE error to bins with high statistical fluctuations.

For monochromatic spectra, the mean values and the standard deviations from gaussian fits of the peak of the count distributions were evaluated.

\section{Results}
\label{results}

\subsection{Simulation studies}
\label{results:simulation}

The response of a CdTe detector to a polychromatic spectrum from an X-ray tube at 120 kVp with W target and 1.6 mm Al filter
was used to determine the elements of the detector response matrix. A detector with pixel size 200 \textmu m was simulated,
a random noise of 1 keV rms was added to the raw energy of each pixel, the width of each energy bin was set to 1 keV
and a noise discrimination threshold of 8 keV was applied to the raw counts and to the analog sums used for the
ACS and the CBRM matrix determination.

\begin{figure}[h]
	\centering
	\includegraphics[width=0.95\textwidth]{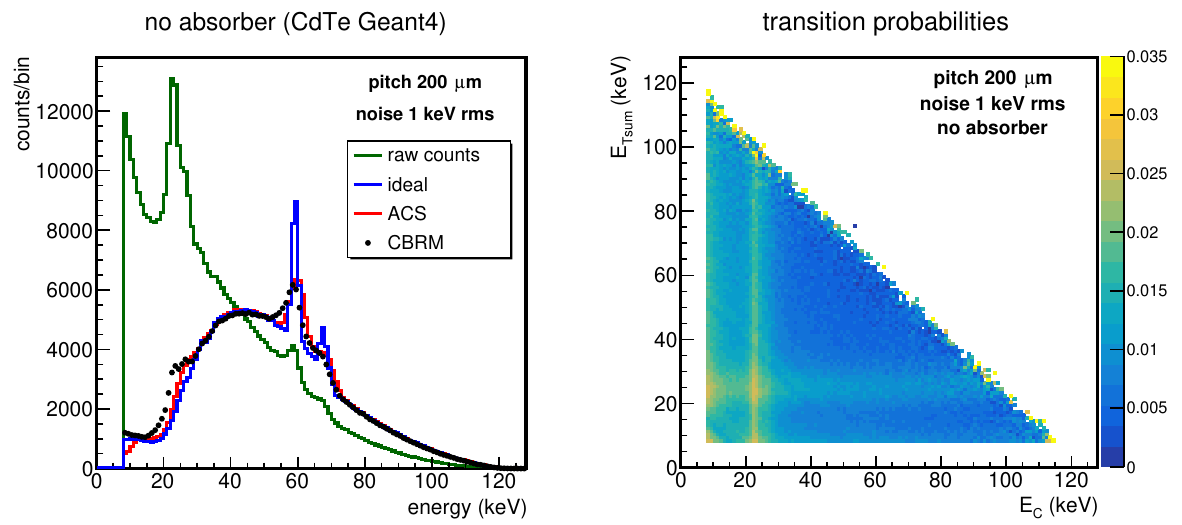}
	\caption{Left: raw count distribution (green histogram), ideal count distribution (blue), result of an ACS algorithm (red)
	and of the CBRM matrix correction (back markers) for a 120 kVp polychromatic spectrum on a CdTe detector of 1 mm thickness
	and 200 \textmu m pixel size (Geant4 simulation). A random noise of 1 keV rms was added to the raw energy of each pixel,
	and a noise discrimination threshold of 8 keV was applied. The CBRM correction was based on a response matrix
	determined with the same data. Right: transition probabilities determined from the raw pixel counts and coincidences.}
	\label{fig6}
\end{figure}

Fig.\ref{fig6}(left) shows the ideal count distribution as a blue histogram, the raw counts as a green histogram,
the results of the ACS algorithm in red, while the black markers correspond to the counts in each energy bin corrected by solving
Eq.\ref{eq8}. Here and in the following figures, the error bars could be hidden by the line width or marker size.
The peak in the raw counts at 23 keV is produced by fluorescence photons from the Cd K-line emitted in other pixels
and absorbed in the pixel under study. The ACS and CBRM methods compensate reasonably well the charge sharing spectrum distortions,
excluding the lower energy region and the fluorescence peak. Both the correction methods are indeed based on charge summing or coincidences between a
limited number of pixels and are not able to compensate for the fraction of fluorescence photons emitted too far and reaborbed in the pixel under study.
The comparison of the restored count distributions with the reference gives MAPE values of 8.0\% (ACS) and 12.0\% (CBRM).
The discrepancy with respect to the ideal distributions are mainly due to the overestimation of the number of events (respectively by 1.8\% (ACS) and 2.5\% (CBRM)),
especially in the low energy bins and around the fluorescence peak. The overestimation is due to the sensitivity
of the charge summing to the noise for the ACS, while for the CBRM to the lack of knowledge of the low energy elements of the response matrix,
excluded by the noise discrimination cut.

Fig.\ref{fig6}(right) shows the transition probabilities determined with Eq.\ref{eq7} using the coincidence counts obtained
from the spectrum of Fig.\ref{fig6}(left) after solving Eq.\ref{eq8} to determine the number of input events.
It is evident that the transition probabilities are particularly high at low energies,
where charge sharing effects are more severe, and in the presence of fluorescence photons
(horizontal and vertical bands at 23 keV).
It is worth noting that the width of the energy bins used to define the response matrix are comparable with the noise rms: this produces a
blurring of the matrix contents but does not significantly affect the restoring capabilities
(the MAPE from the comparison of the CBRM corrected counts with the ideal spectrum for a noiseless detector is only slightly reduced
from 12.0\% to 10.8\%).

\begin{figure}[ht]
	\centering
	\includegraphics[width=0.48\textwidth]{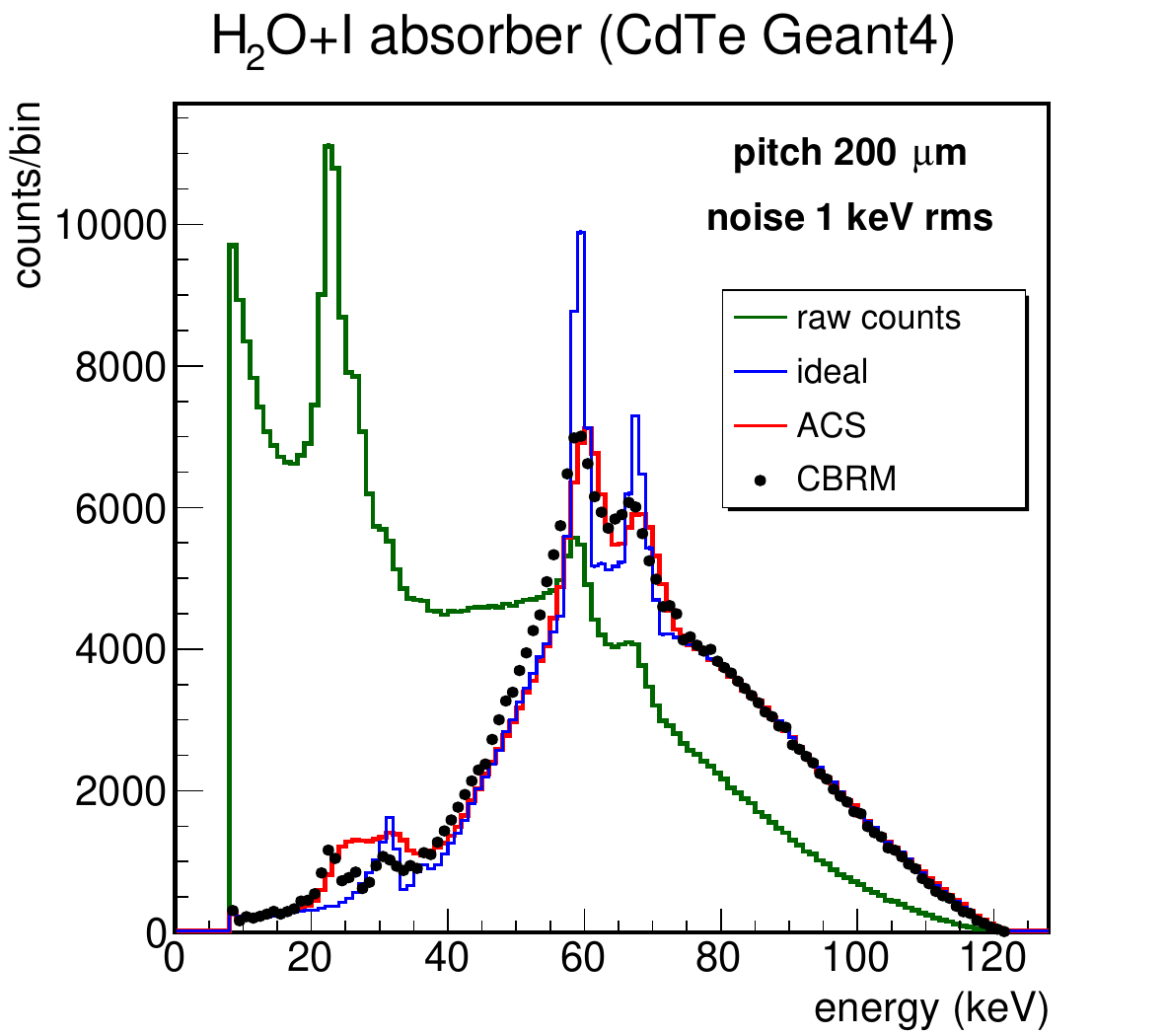}
	\caption{Raw count distribution (green histogram),  ideal count distribution without CS (blue histogram),
	results from an ACS algorithm (red histogram) and CBRM correction (back markers) for a 120 kVp polychromatic spectrum attenuated
	by 120 mm of H2O and 0.2 mm of Iodine. A CdTe detector of 1 mm thickness and 200 \textmu m pixel size was simulated with Geant4.
	A random noise of 1 keV rms and a noise rejection threshold of 8 keV were applied.
	The CBRM correction used the response matrix determined with data collected without attenuation.}
	\label{fig7}
\end{figure}

The same response matrix determined above was applied to a simulation of a 120 kVp spectrum attenuated by 100 mm of H$_{2}$O and a 0.2 mm of Iodine, with the results shown in Fig.\ref{fig7}.
In these conditions the MAPE values from the comparison of the corrections with the ideal count distribution are 18.8\% (ACS) and 20.9\% (CBRM).
The total number of counts are overestimated by 2.5\% (ACS) and 5.2\% (CBRM). Worse results were obtained if the CBRM correction
was performed with a matrix determined with the count and coincidences collected with the same attenuated input (MAPE=28.9\%).
Even in other conditions (different pixel size, noise discrimination threshold, noise level), the application of a correction matrix
determined with a non-attenuated spectrum gives in general better results than using a matrix determined with the same input data,
due to the lack of information at low energies in the attenuated spectrum.
These observations demonstrate the possibility to apply the response matrix determined with a calibration acquisition to data acquired
with other input spectra attenuated by materials placed between the source and the detector.

\begin{figure}[ht]
	\centering
	\includegraphics[width=0.95\textwidth]{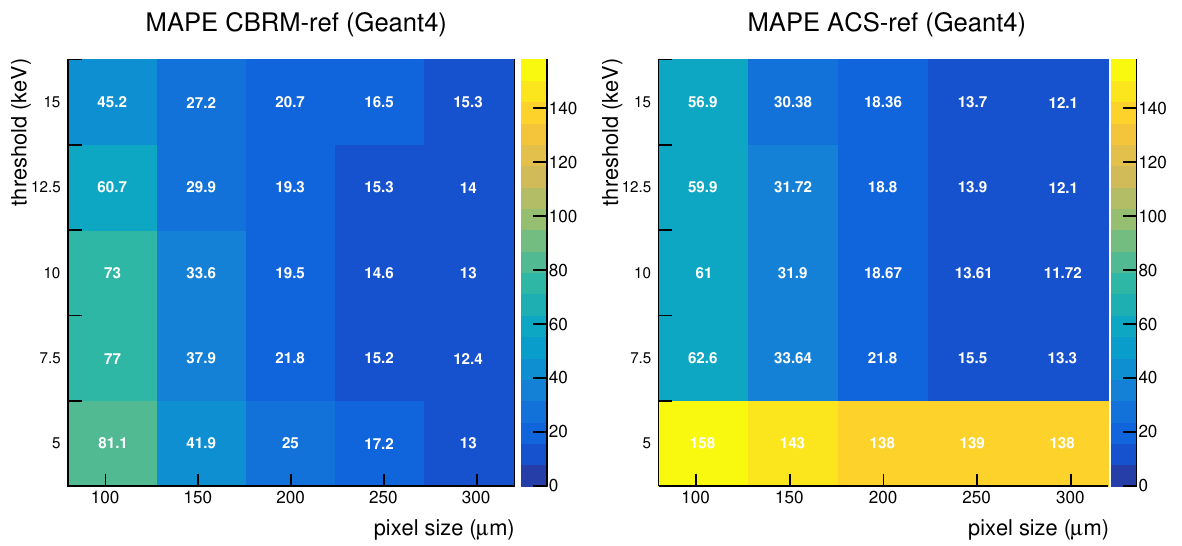}
	\caption{MAPE values from the comparison of the CBRM correction (left) and the ACS correction (right) with the reference spectrum
	of an ideal detector, as a function of the pixel size and the noise discrimination threshold. The input spectrum from a 120 kV X-ray tube was
	attenuated by 120 mm of H$_{2}$O and 0.2 mm of Iodine, and a random noise of 1 keV rms was added to the output of each pixel.
	The results were obtained from Geant4 simulations.}
	\label{fig8}
\end{figure}

The performance of the ACS and CBRM corrections depends on the pixel size and on the level of the noise discrimination threshold.
Fig.\ref{fig8} shows the MAPE values from the comparison of the CBRM (left) and ACS (right) count distributions with the reference one,
as a function of the pixel size and of the noise discrimination threshold, for a fixed noise level of 1 keV rms.
The distributions refer to the same attenuated spectrum of Fig.\ref{fig7}, and the response matrix was determined with a non-attenuated spectrum
applying the same noise discrimination cut used for the correction. At low threshold levels, the ACS count distributions have very high MAPE values.
This is due to the enhanced sensitivity to noise of the analog sums employed in the ACS algorithm. Fake counts appear from summing nodes exceeding the noise discrimination
threshold due to the sum in quadrature of the noise contributions from 4 pixels. Even the CBRM determination makes use of analog sum of the
energies in 8 pixels, but these are used in coincidence with a hit in the reference pixel above the noise discrimination threshold to determine the response matrix.
The correction is afterwards applied to the number of counts from single pixels, which for a given threshold are less sensitive to noise effects than the analog sum
employed by ACS methods. In general, to avoid fake counts due to noise, the ACS requires a higher minimum threshold with respect to the CBRM correction.
For a fixed pixel size, the MAPE for the ACS is stable as a function of the threshold once the noise discrimination threshold is high enough to filter fake counts.
For pixel size of 200 \textmu m or more, the MAPE values from CBRM are sligtly worse than ACS at high thresholds, due to the reduced size of the response matrix used in Eq.\ref{eq9c} to restore the count
distribution. 
For pixel size lower than 200 \textmu m and noise discrimination threshold greater than 5 keV, the performance of the CBRM and ACS corrections are similar.
For the simulated 1 mm thick CdTe detector, both the correction methods have poor performance for pixel size of 100 \textmu m or less.

The effectiveness of the CBRM correction depends also on the number of energy bins used for the definition of the response matrix.
In general, a high number of energy bins is desirable for an accurate determination of the matrix; this can be obtained by changing the
thresholds of the two discriminators of Fig.\ref{fig1}(right) in small steps. However, the typical number of comparators and the number of
energy bins available in standard pulse-height acquisition systems is limited. The procedure described in Section\ref{mehod:model:matrix_reduction}
can be followed to reduce the response matrix to match the number of energy bins of a standard acquisition.

\begin{figure}[h]
	\centering
	\includegraphics[width=0.48\textwidth]{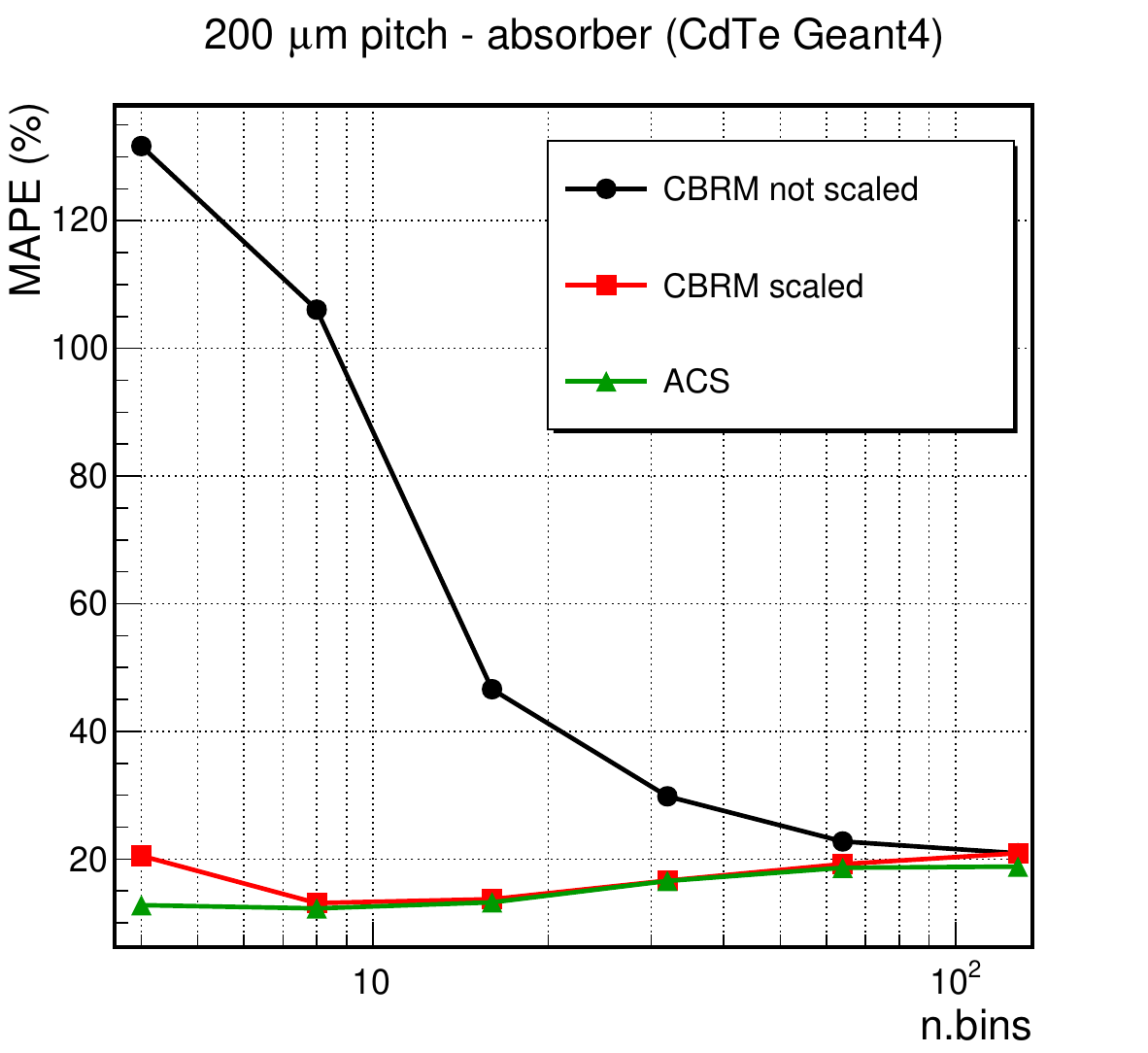}
	\caption{MAPE values from the comparison of CBRM reconstructed spectra with the count distribution from an ideal detector as a function of the number N of energy bins.
	A Geant4 simulation of a 1 mm thick CdTe sensor with pixel size 200 \textmu m was used. A random noise of 1 keV rms and a noise discrimination threshold of 6 keV were applied to the output
	of each pixel. The input consisted of 120 kVp X-rays attenuated by 100 mm of H$_{2}$O and 0.2 mm of Iodine.
	Black markers: CBRM reconstuction with a response matrix determined with a non-attenuated spectrum employing NxN energy bins.
	Red markers: CBRM reconstruction with a response matrix determined with 128x128 bins and rescaled to NxN bins. Green markers: reconstruction with an ACS algorithm.}
	\label{fig9}
\end{figure}

Fig.\ref{fig9} shows the MAPE values from the comparison of CBRM reconstructed count distribution with the spectrum of an ideal detector,
as a function of the number of bins used in the acquisition. The Geant4 simulation comprised a 1 mm thick CdTe sensor with 200 \textmu m pixel,
a noise rms of 1 keV and a noise discrimination threshold of 6 keV. The MAPE values refer to the reconstruction of a 120 kVp spectrum
attenuated with 100 mm of H$_{2}$O and 0.2 mm of Iodine (Fig.\ref{fig7}) with the CBRM obtained with the non-attenuated spectrum of Fig.\ref{fig6}(left).
The black markers in Fig.\ref{fig9} correspond to the MAPE values determined by applying the CBRM matrix with the same number of bins of the acquisition
of the attenuated spectrum, shown in the horizontal axis in figure. The red markers in figure are the MAPE value for a CBRM reconstruction based
on a response matrix with 128x128 energy bins (bin width 1 keV) and rescaled to the number of bins used in the acquisition
of the attenuated spectrum, shown in the horizontal axis. The green markers are the MAPE values from the comparison of the ACS algorithm with the reference spectrum.
The performance of the CBRM reconstruction worsens by reducing the number of energy bins for the matrix determination, while the use of a
rescaled matrix allows to provides a good spectrum restoration independently of the number of energy bins used for the acquisition of the
attenuated spectrum. The results from the rescaled matrix are comparable with those obtained with the ACS algorithm.

\subsection{Study of experimental data}
\label{results:timepix}

The analysis of the data collected at the SYRMEP beamline with the silicon detector and the Timepix4 ASIC was performed using energy bins of 0.5 keV width.
The response matrix was determined by merging all the data collected at 18 different energies,
to define the splitting probabilities over a wide energy range. The CBRM matrix was therefore applied to recover
the spectrum of each single acquisition, and the results were compared with the spectra obtained from a
clustering algorithm based on 3x3 pixel blocks (Sect.\ref{method:validation}).

\begin{figure}[h]
	\centering
	\includegraphics[width=0.95\textwidth]{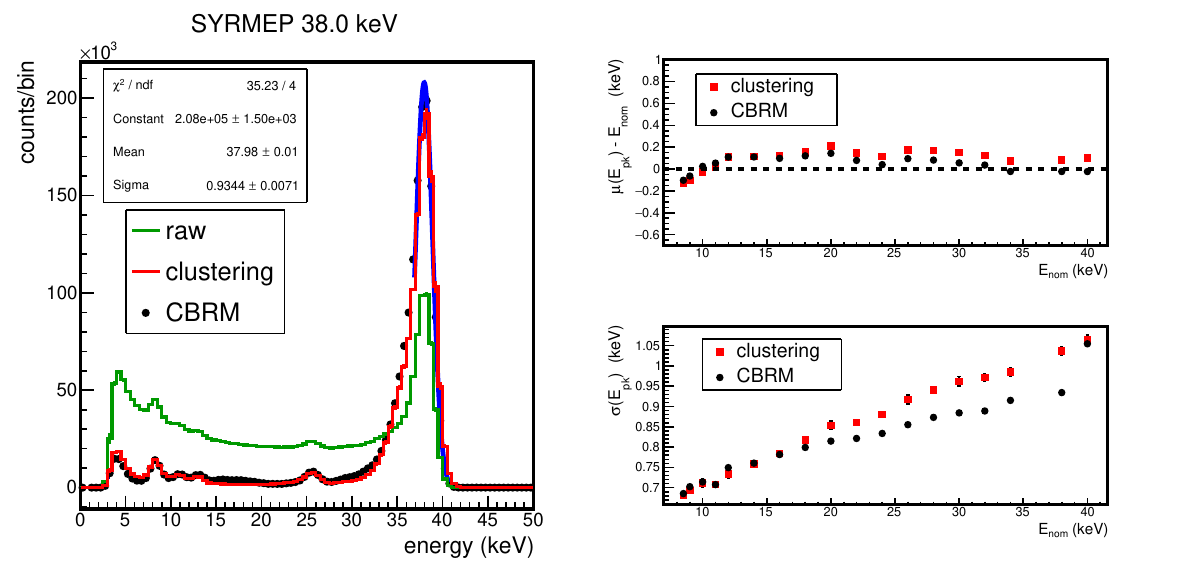}
	\caption{Left: Pulse-height distribution of raw counts (green), of counts from a 3x3 clustering algorithm (red),
	and from the CBRM correction (black marker) for monochromatic 38 keV X-rays (SYRMEP beamline at the Elettra synchrotron)
	acquired with a 300 \textmu m thick Si detector readout by a Timepix4 ASIC. The result of a gaussian fit to the peak
	is shown as a blue curve. Right: mean and sigma values from gaussian fits of the peak of the count distributions as a function
	of the nominal X-ray energy for the clustering (red dots) and CBRM correction (black dots).}
	\label{fig10}
\end{figure}

Fig.\ref{fig10}(left) shows the count distributions for an irradiation with monochromatic
38 keV X-rays: the green histogram corresponds to the raw counts, the red to the count distribution
from the clustering algorithm, the black markers to the CBRM correction.
 The MAPE from the comparison of the CBRM correction with the clustering is 28\%. At this energy the total number of counts from the CBRM correction is 3.7\% higher than the
number of clusters. A gaussian fit to the peak of the CBRM distributions (excluding the tail at lower energy) is also shown
in figure as a blue curve.

Gaussian fits to the peaks of the clustering and CBRM distributions were performed separately for each of the 18 energies between 8.5 and 40 keV used in the acquisitions.
The plot at the top of Fig.\ref{fig10}(right) shows the difference between the mean values of the gaussian fits $\mu(E_{pk})$ and the nominal energy $E_{nom}$
as a function of the nominal energy for the clustering (red dots) and the CBRM correction (black).
The plot at the bottom shows the standard deviations from the fits.

The CBRM reproduces well the results from the clustering algorithm, with typical MAPE values from the comparison of the two distributions
between 10\% at low energies and 30\% at higher energies. The discrepancies are mainly due to an offset of about 0.1 keV between the peak positions of the two distributions for $E_{nom}>20$ keV and to an excess of counts
from the CBRM correction up to 3\% at the highest energies.
Excluding the lower energies and the anomalous value at 40 keV,
the standard deviations for the CBRM corrections are smaller than those obtained after clustering.

\begin{figure}[h]
\centering
	\includegraphics[width=0.95\textwidth]{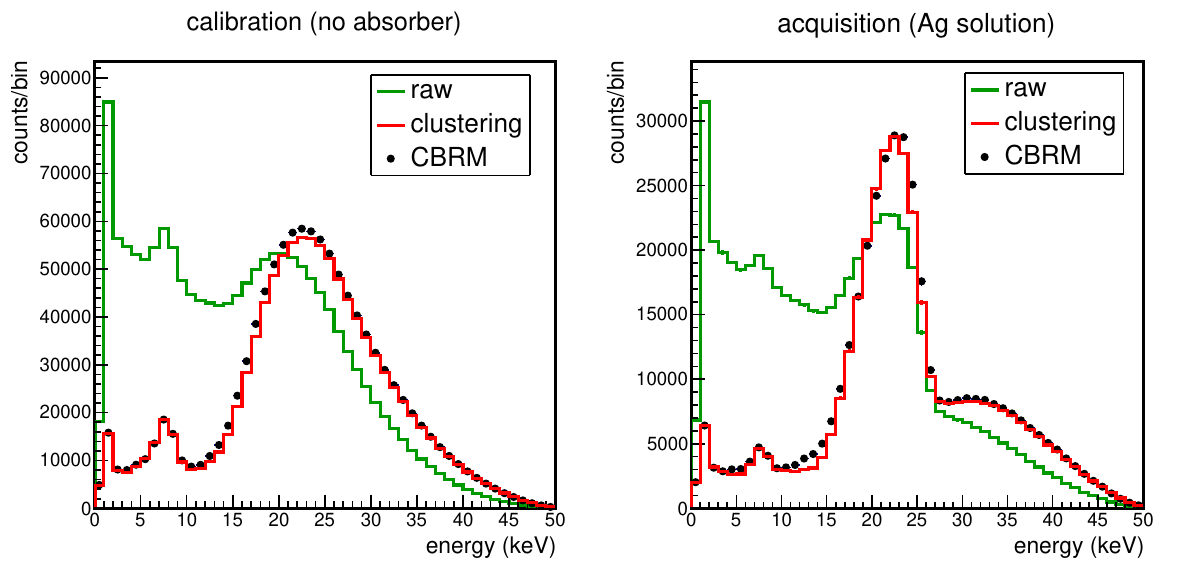}
	\caption{Pulse-height distribution of raw counts (green), of counts from a 3x3 clustering algorithm (red), and from the CBRM correction (black marker)
	for an acquisitions of a 50 kVp X-ray beam without absorber (left) and with the beam attenuated by an Ag solution. A ROI of 1500 pixels was selected
	and the CBRM correction was based on the matrix determined with the spectrum without absorber.}
	\label{fig11}
\end{figure}

For the acquisitions with a polychromatic 50 kVp X-ray beam, the response matrix was determined with a flat-field illumination
without absorber and applied to restore the count distribution of a spectrum attenuated by an Ag solution.
The count distributions for the two spectra are reported in Fig.\ref{fig11}, respectively for the acquisition without absorber (left) and
with the absorber (right). The green histograms correspond to the raw counts, the red to the results from a 3x3 clustering algorithm,
the black markers to the counts corrected with the CBRM.
The MAPE error from the comparison between the clustering and corrected distributions are 3.6\% (without absorber) and 6.2\% (with absorber).
In both cases the CBRM correction provides a total number of counts 3.4\% greater than the number of reconstructed clusters.

The CBRM matrix determined with a reference spectrum provides a restored spectrum similar to that from a clustering
algorithm. The dip at the Ag K-edge of 25.5 keV is clearly visible in both reconstructed spectra.
No attempt was done in this study to quantify the concentration of Ag in the solution with a
K-edge analysis. However, the good comparison with the performance of a standard clustering algorithm indicates that the
CBRM reconstruction is able to restore the charge sharing spectral distortion and could be applied for contrast agent
identification and quantitative material discrimination.

\section{Discussion}
\label{discussion}

The studies based on simulations and on X-ray data collected with a Timepix4 hybrid silicon detector showed the
effectiveness of the CBRM method to restore the spectral distortions due to charge sharing effects. In the simulation studies
the spectra restored with the CBRM and a standard ACS algorithm were compared with the output of an ideal detector of infinite pixel size.
For the experimental data, an offline 3x3 clustering algorithm was used for comparison.
In both cases the results from the CBRM corrections were close to those obtained by analog summing or offline clustering.
It was also demonstrated that the application of the CBRM matrix is independent of the acquired spectra.

In the simulation the ACS resulted to be more sensitive to noise than CBRM when the minimum energy threshold is close to the noise level.
This effect was not visible in the experimental tests because the clustering was performed after applying a noise discrimination
threshold to each of the detector pixels, while in a charge summing algorithm uncorrelated noise contributions
from the pixels of each summing node are summed in quadrature before the application of a noise cut. Even the determination of the CBRM matrix is based on the analog
sum of the signals of 8 pixels, but the correction is less sensitive to the noise at low energy, because it is applied to pixel counts above the minimum threshold.

The CBRM performance degrades at higher noise discrimination thresholds. This is because the CBRM algorithm requires that the
charge sharing transition probabilities are determined over a wide range of energy values, ideally from 0 to the maximum
energy of the spectrum to be recovered. The noise discrimination threshold prevents the measurement of the transition probabilities
for low energy bins. In particular, the term $q_{i,0}$ in the denominator of Eq.\ref{eq9c} cannot be determined with accuracy or
is not known at all and set to 0. The effect is an overestimation of the number of counts, dependent on the noise discrimination threshold.
The MAPE errors reported for the comparison of the CBRM corrections with other reference spectra are mainly due
to an excess of a few per cent of the total number of counts from the CBRM method.

The performance of both CBRM correction and charge summing algorithms degrades for small pixel size, when charge sharing
involves clusters with more than 2x2 pixels. In the simulation of a 1 mm thick CdTe sensor, for pixel size below 100 \textmu m
the CBRM and ACS algorithms fail in recovering of the input spectrum.

Another assumption of the CBRM model is that the ideal count distribution inside each energy bin is uniform,
such that the transition probabilities are independent of the energy distribution within each energy bin.
The model is also based on the hypothesis that the bin index corresponding to the input energy is the sum of the bin
indices of the reference and peripheral pixels. The validity of these assumptions requires energy bins of small width.
In the acquisition scheme of Fig.\ref{fig1} for the determination of the CBRM matrix, the number of bins
depends only on the settings of two DAC converters and the energy range can therefore be divided in a high number of
energy bins whose widths are small enough to be close to these hypotheses.
However, the number of thresholds used in standard multi-discriminator electronics is in general lower than the number
of thresholds needed for an accurate determination of the response matrix. It was shown that it is possible to reduce
the response matrix to match the number of energy bins of a standard acquisition system.
The correction in such a case is almost insensitive to the number of bins employed by the standard
pulse-height acquisition system and is more accurate than using a response matrix defined with the same energy bins of the standard acquisitions.

There are some limitations in the present study. The simulation was very simple, assuming a fixed charge cloud size independent of the photon energy and
of the interaction depth, neglecting charge trapping, polarization, and inter-pixel gaps.
However, the method does not rely on an accurate simulation of charge sharing effects, as all the relevant information needed to restore the
spectral information are determined by correlation measurements in the calibration phase.

The assessment of the method was performed only with flat-field irradiations, where the spectra and the intensities in all the PCD pixels were identical.
This is not the case in imaging applications, where the energy count distributions in various pixels are in general different.
In principle, with some approximations, the CBRM matrix could be used to unfold the spatial dependence of the spectral information,
but this possibility was not investigated in this paper. The deconvolution of the spatial information using a spectral point spread function based on the CBRM matrix and its
performance in terms of spatial resolution and spectral restoring is postponed to future studies.

This study also neglects pileup effects at high rates which could mimic correlated counts in two neighbouring pixels
providing a wrong prediction of the response function. The problem could be limited in the calibration phase by using
low beam currents. However, in a standard acquisition, especially for clinical imaging, pileup effects are not negligible and a combined
spectral correction for CS and PU distortions is desirable. In \citep{MONACO2025170276} a model-independent method to restore pileup spectral
distortions by using two acquisitions with different input rates was proposed. In the future the possibility to factorize this
 PU correction and the CS correction with a CBRM matrix will be explored.

 The model is also based on the measurement of counts in equidistant energy bins. Possible extensions of the method to energy bins
 of different widths are possible and will be studied in the future.

There are already many proposals to mitigate charge sharing and/or pileup effects with a post-processing of the recorded counts,
but they are based on models, simulations or data fitting which are specific for a given detector or acquisition setting
and in general are not applicable in a generic situation. Model-independent method proposed for charge sharing mitigation are
based on dedicated readout electronics, like anticoincidence circuits \citep{Shankar2018}, the already mentioned
analog charge summing techniques, and multi-energy inter-pixel coincidence counters (MEICC) \citep{Taguchi2020}.
The implementation of these solutions is more complex than CBRM, requiring multiple inter-pixel charge summing and communications or
many coincidence gates for MEICC.

The CBRM method has some similarities with the MEICC technique proposed to collect information on how charge sharing
is distributed between different energy bins of neighbouring pixels using coincidence counts.
However, there is an important difference: in the MEICC method the number of counts for the AND logical correlations are incremented
whenever any of the neighbouring pixels measures an energy within a given energy window. In practice, the pixels
around a reference one are considered as independent in terms of charge sharing probabilities.
In our opinion, this approach is valid only for events with charge sharing involving two pixels.
The probability of charge sharing for events of multiplicity greater than two cannot be considered as a simple
combination of the individual sharing probabilities for pairs of involved pixels, but can be evaluated
only by considering all the pixels around the reference one as a whole, using the sum of their energies for the coincidence counts.
In addition, our proposal does not require the complex network of AND gates foreseen by the MEICC technique for all the detector pixels,
but only a simple AND logic between one pixel and the sum of the signals of the 8 neighbours to build the response function in a calibration stage.
The calibration can be performed with a dedicated small detector with the same sensor employed for subsequent imaging acquisition, or implementing the
coincidence readout in a small part of the full size detector.
The following acquisitions are performed with standard multiple discriminators and counting electronics, with no modification to the existing X-ray PCDs.
In both the MEICC method and the proposed CBRM method, no additional dead-time is added during the acquisition, allowing higher acquisition rates
with respect to readouts with ACS circuits.

\section{Conclusions}
\label{conclusions}

A method is proposed for the offline compensation of charge sharing spectral distortions in photon counting X-ray acquisitions.
The method is based on the characterization of charge sharing effects for a particular detector through a response matrix
(named CBRM) whose elements are determined by counting the coincidences between the signals from one pixel and
from the analog sum of its 8 neighbours for different combinations of energy bins.
Once the CBRM matrix is determined with a polychromatic spectrum, it can be applied to restore charge sharing spectral
distortions of other spectra using as input only the counts above different thresholds of a conventional
multi-discriminator readout electronics.

The CBRM method was assessed with a Geant4 simulation of a 120 kVp X-ray spectrum, by comparing the corrections with a
reference spectrum of an ideal detector with infinite pixel size and with the results of a standard analog charge summing technique.
Charge sharing spectral distortions and the effectiveness of their corrections depend on many factors, as the photon energy,
the pixel size, the noise level, the noise discrimination threshold. In most of the simulated conditions, a reasonable agreement
of the CBRM corrections with the reference spectrum is achieved, with results very similar to those obtained with an ACS method.
For example, for a 120 kVp attenuated spectrum, a pixel size of 200 \textmu m, a noise level of 1 keV rms, and a  minimum threshold of 10 keV,
the MAPE errors from the comparison of the corrections with the reference spectrum are 19.5\% for the CBRM method and 18.7\% for ACS.
At low noise discrimination threshold the ACS method is more sensitive to noise than the CBRM correction.

The experimental validation of the method was performed with data of monochromatic and polychromatic X-ray beams
collected with a hybrid silicon detector readout by the Timepix4 electronics. The CBRM corrections were compared with the results of an offline
clustering algorithm, providing similar results in restoring the spectra
(as an example the MAPE error from the comparison of the two spectra for 50 kVp X-ray beam was 3.8\%).
Gaussian fits of the reconstructed peaks of monochromatic spectra from the two corrections provide similar mean and standard deviation values.

It is also demonstrated that a CBRM matrix determined with a fine division of the energy range can be adapted to a lower number of bins employed
in a multi-discriminator acquisition without affecting the accuracy of the spectrum reconstruction.

In conclusions, the proposed method is independent of models or parameterizations, is simple to realize, as
the required coincidence electronics involve only a block of 9 pixels in a calibration run, and the response matrix can be applied to correct different spectra collected by
a standard PCD with pulse-height capability without introducing additional dead-times.

\section{Acknowledgments}

This work was supported by the MEDIPIX4 project
funded by the INFN (National Institute for Nuclear Physics), CSN5
(National Scientific Commission 5). We thank the Medipix collaboration at CERN for the support
in the operation of the Timepix4 electronics.

\section{Declaration of competing interest}

The authors declare that there are no known conflicts of interest
associated with this publication.

\bibliographystyle{unsrtnat}
\bibliography{arxiv_v1}  

\end{document}